\documentclass{article}
\usepackage[T1]{fontenc} 
\usepackage[utf8]{inputenc} 
\usepackage{ismir,amsmath,cite,url}
\usepackage{graphicx}
\usepackage{float}
\usepackage{color}
\usepackage{booktabs}
\usepackage{caption}
\usepackage{environ}         
\usepackage{etoolbox}        
\usepackage{amsfonts}

\newlength{\myl}
\let\origequation=\equation
\let\origendequation=\endequation

\RenewEnviron{equation}{
  \settowidth{\myl}{$\BODY$}                       
  \origequation
  \ifdimcomp{\the\linewidth}{>}{\the\myl}
  {\ensuremath{\BODY}}                             
  {\resizebox{\linewidth}{!}{\ensuremath{\BODY}}}  
  \origendequation
}

\usepackage[firstpage]{draftwatermark}
\definecolor{lightgray}{rgb}{0.9,0.9,0.9}
\definecolor{darkgray}{rgb}{0.4,0.4,0.4}
\SetWatermarkFontSize{12pt}
\SetWatermarkScale{1.1}
\SetWatermarkAngle{90}
\SetWatermarkHorCenter{202mm}
\SetWatermarkVerCenter{170mm}
\SetWatermarkColor{darkgray}
\SetWatermarkText{Late-Breaking / Demo Session Extended Abstract, ISMIR 2024 Conference}

\def\lbd{}	

\title{Self-supervised Multi-view Learning for Disentangled Music Audio Representations}

\oneauthor
{Julia Wilkins \hspace{1cm} Sivan Ding \hspace{1cm} Magdalena Fuentes \hspace{1cm} Juan Pablo Bello  \vspace{0.2cm}}
{Music and Audio Research Lab, New York University \\ 
{\tt 
\{jw3596,sivan.d,mfuentes,jpbello\}@nyu.edu} \\
{\tt{\footnotesize \url{https://juliawilkins.github.io/marlbymarl/}}}}

\def\authorname{J. Wilkins, S. Ding, M. Fuentes, and J. P. Bello}

\usepackage[bookmarks=false,pdfauthor={\authorname},pdfsubject={\papersubject},hidelinks]{hyperref}

\newcommand{\insertfig}{\includegraphics[width=\textwidth,trim={0.5cm 1.1cm 0cm 1.4cm}, height=5.2cm]{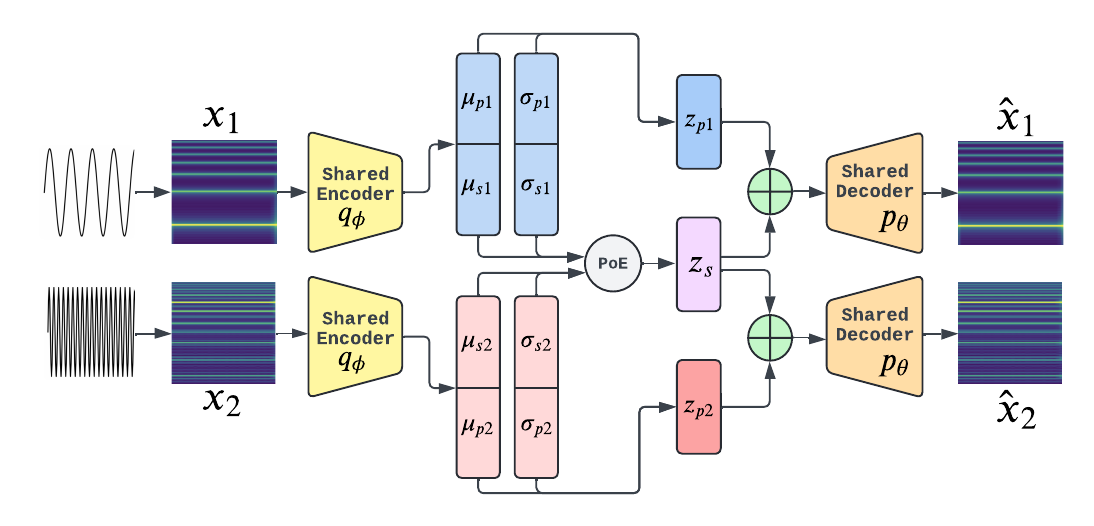}
\captionsetup{type=figure}
\setlength{\belowcaptionskip}{7pt} 
  \captionof{figure}{
    Our multi-view audio representation learning method. Pairs of audio spectrograms with common timbre but unique frequency parameters are passed to the disentanglement model.
  }
  \label{fig:block_diag}}

\makeatletter
\apptocmd{\@maketitle}{\centering\insertfig}{}{}
\makeatother

\begin{document}
\pdfoutput=1




\maketitle


\sloppy

\vspace{-0.3cm}
\begin{abstract}
\vspace{-0.165cm}
Self-supervised learning (SSL) offers a powerful way to learn robust, generalizable representations without labeled data. In music, where labeled data is scarce, existing SSL methods typically use generated supervision and multi-view redundancy to create pretext tasks. However, these approaches often produce entangled representations and lose view-specific information. We propose a novel self-supervised multi-view learning framework for audio designed to incentivize separation between private and shared representation spaces. A case study on audio disentanglement in a controlled setting demonstrates the effectiveness of our method.
\end{abstract}

\vspace{-0.5cm}

\section{Introduction}\label{sec:introduction}


SSL uses pretext tasks to uncover patterns from unlabeled data. In single-view SSL, a model learns from one perspective of the input via information restoration \cite{Balestriero2023ACO}. Multi-view SSL, however, utilizes distinct views 
to generate supervision, assuming shared information across views suffices for downstream tasks \cite{liang2024factorized}. These methods align and contrast information across views for learning.

Recent multi-view SSL studies use contrastive learning to treat audio segments or augmentations as transformed views \cite{saeed2021contrastive, spijkervet2021contrastive}. However, these approaches neglect the intrinsic structure of music audio, entangling representations with attributes like timbre, frequency, and  tude. They also focus on shared information across views, missing task-relevant, view-specific details.

Previous research has primarily focused on separating pitch and timbre to encode disentangled music representations, often designing dedicated encoders for each attribute. Some approaches train generative models with explicit supervision on pitch and timbre latents \cite{DBLP:conf/ismir/LuoAH19}, while others reduce supervision by applying auxiliary metric-based regularization in the latent spaces \cite{luo2020unsupervised, tanaka2021pitch, luo2022robustunsuperviseddisentanglementsequential, 10447564}.

We propose a novel self-supervised multi-view learning framework for music audio, inspired by multi-view/multimodal disentanglement \cite{lee2020private, multivae}, which explicitly separates shared and private representations. This approach preserves the uniqueness of each view while capturing common latent factors. As a case study, we tackle music audio disentanglement from a multi-view learning perspective with self-supervision. We validate our method using Syntone \cite{brima2024syntone}, a dataset with controlled variations in music attributes such as timbre and frequency.

\vspace{-0.3cm}

\section{Method}

Our method is shown in Figure \ref{fig:block_diag} for the case of two views. The model receives as input a pair of normalized log mel spectrograms, denoted as $(x_1, x_2)$, which are characterized by identical timbre (waveform class) but distinct frequency. 
Factorized latents $z_1 \sim q_{\phi}(z|x_1)$ and $z_2 \sim q_{\phi}(z|x_2)$ are inferred from a shared encoder $q$ parameterized by $\phi$. To separate the private and shared spaces, we split the latents into $z_1=[z_{p1}, z_{s1}]$ and $z_2=[z_{p2}, z_{s2}]$, where $z_{p1}, z_{p2} \in \mathbb{R}^{D_p}$ are the private latents of each view and $z_{s1}, z_{s2} \in \mathbb{R}^{D_s}$ correspond to the private-shared latents. 
Following \cite{hinton2002training}, we use a PoE-based consistency model to approximate the true posterior $p(z_s|x_1, x_2)$ so that we effectively make $z_{s1}=z_{s2}=z_s$. For all the continuous latent variable $\bm z$, we assume $p(\mathbf{z})=\mathcal{N}(0, I)$.

We set the latent dimensions to be $D_{p}=D_{s}=8$ and sample from the parameterized distributions to arrive at three $8$-dimensional latent variables: $\mathbf{z_{p1}}$, $\mathbf{z_{p2}}$, and $\mathbf{z_{s}}$ corresponding to the $x_1$-private, $x_2$-private, and shared representations, respectively. We concatenate each private latent $z_{pi, i\in \{1,2\}}$ with the shared latent $z_s$ as input to the shared decoder $p_\theta$ parameterized by $\theta$.

Following DMVAE \cite{lee2020private} and $\beta$-TCVAE \cite{chen2018isolating}, the tractable evidence lower bound of our model assumes the form for each sample pair $\bm x$ and $N$ views of the data:

\vspace{-0.55cm}
\begin{align}
\label{eq:loss}
\begin{split}
\sum_i^N \mathbb{E}_{p(x_i)}[& \mathbb{E}_{q_\phi(z_{p_i} |x_i), q_\phi(z_s, |\boldsymbol{x})}[\log p_\theta(x_i | z_{p_i}, z_s)] \\
&-K L(q_\phi(z_{p_i} \mid x_i) \| p_\theta(z_{p_i}))\\
&-K L(q_\phi(z_s \mid \boldsymbol{x}) \| p_\theta(z_s))].
\end{split}
\end{align}
The objective to maximize is the sum of the reconstruction accuracy compensated by the KL-Divergence for each view $i$. Particularly, each KL term in Eq (\ref{eq:loss}) is decomposed into mutual information $MI(x_i;z)$, total correlation $TC(z;\prod_kq_{\phi}(z_k))$, and dimension-wise KL $DKL(q_{\phi}, p)$, with $\alpha$, $\beta$, and $\gamma$ as penalty weights respectively. 



\vspace{-0.5cm}

\section{Experimental Design}

\textbf{Data Synthesis:} We generate a controlled dataset of $32,000$ 1-second audio samples and corresponding normalized log-mel spectrograms with varying timbre and frequency factors, similar to the data generation of SynTone \cite{brima2024syntone}. We randomly choose \textbf{pairs} of log-mel spectrograms with the \textit{same timbre} but with \textit{different frequencies} to incentivize our model to learn shared timbre information and varying frequency information.

\noindent\textbf{Model Training:} We train our model for 100 epochs with a learning rate of $0.001$ using the Adam optimizer. We use $\alpha=\beta=0$ and $\gamma=0.1$ for weighting the decomposed KL-Divergence in Eq. \ref{eq:loss} (explained in Sec. 4).


\noindent\textbf{Evaluation Design:} We use mutual information matrices between the latents and the factors to evaluate how much information about each factor is contained in each subspace and within the latent dimensions. For downstream classification, we use our trained disentanglement model as a feature extractor and train a lightweight classifier to predict timbre and quantized frequency. We compare the downstream performance of different latents on both tasks to understand the information contained in each subspace.

\vspace{-0.2cm}
\section{Results and Discussion}\label{sec:exp}

\begin{figure}
\begin{center}
 \includegraphics[alt={ISMIR 2024 LBD template test image},width=0.95\columnwidth, trim={1cm 0.5cm 2cm 1cm},height=6cm]{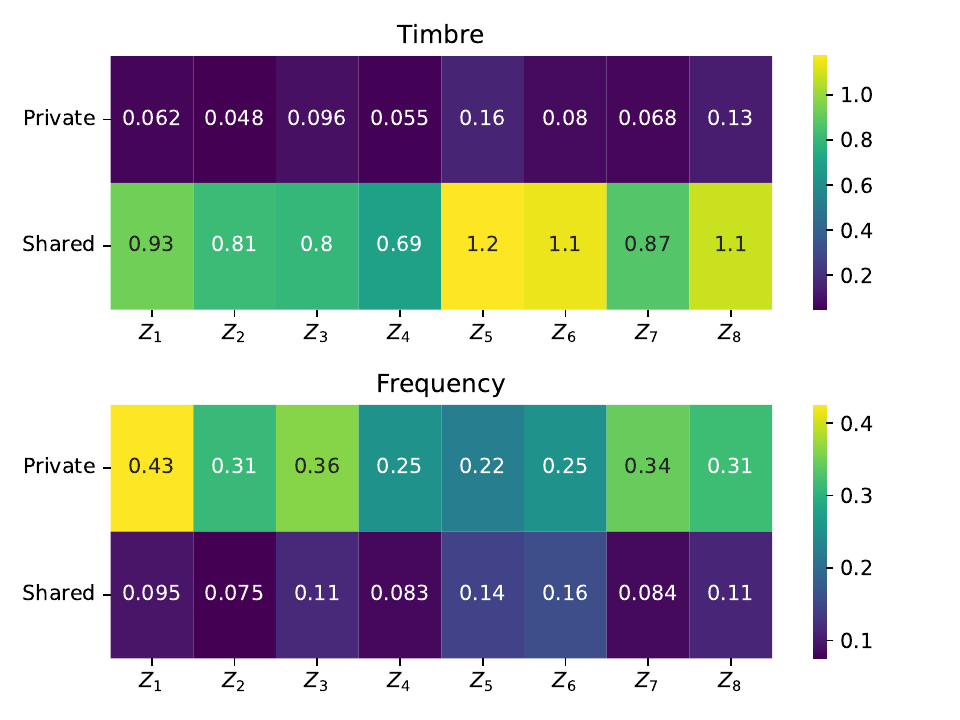}
 \setlength{\belowcaptionskip}{-20pt} 
 \caption{Mutual information between latent dimensions ($z_k$) within each latent and the generative factors.}

 \label{fig:example}
 \end{center}
\end{figure}

In Figure \ref{fig:example}, we show that our model is able to clearly disentangle frequency information into the private subspace and timbre information into the shared in terms of the mutual information between factors and latent dimensions. This observation supports our hypothesis that a multi-view framework combined with a paired dataset that contains shared and private information should be able to learn separate embedding subspaces pertaining to different factors.


Our downstream classification performance in Table \ref{tab:class_values} shows similar trends to the above: when using the private latent or a combination of the private and shared, our model performs very well at predicting frequency, and when using the shared latent or combination of both, we are able to classify timbre successfully. 


\vspace{-0.1cm}
\begin{table}[ht]
\centering
\begin{tabular}{l|c|c}
&  \textbf{Timbre} & \textbf{Frequency} \\
Latent Used  & \textbf{$n_c=4$} & \textbf{$n_c=21$} \\
\hline
\textbf{Private} &  $0.7431$ & $\mathbf{0.7663}$   \\
\textbf{Shared} & $\mathbf{0.9741}$ & $0.2978$ \\
\textbf{Both}    & $\mathbf{0.9809}$ & $\mathbf{0.8772}$ \\
\end{tabular}
\setlength{\belowcaptionskip}{-7pt} 
\caption{Classification accuracy on the test set using different embedding subspaces from our trained encoder.}
\label{tab:class_values}
\end{table}

We also perform an ablation of weights of the decomposed KL-Divergence terms in Eq. \ref{eq:loss} for $\alpha$, $\beta$, and $\gamma$. Surprisingly, we found that higher weights of $\alpha$ (MI) and $\beta$ (TC) negatively affect private-shared disentanglement, while increasing $\gamma$ as a simple regularizer on each latent has the biggest impact on improving performance.

\vspace{-0.3cm}
\section{Conclusion and Future Work}\label{sec:con} 
We present a novel architecture for disentanglement of view-specific and shared latent subspaces in a controlled audio setting. We show that our hypothesis holds in that unique frequency information is encoded strongly into view-specific representations and common timbre information is encoded best in the shared latents. In the future we will experiment with expanding the objective function to further improve subspace-level latent disentanglement, and also plan to apply our method to real music samples.




\bibliography{ISMIRtemplate}

%
%
%
%
%

\end{document}